\documentclass[acmsmall,screen]{acmart}
\usepackage{array}
\usepackage[most]{tcolorbox}
\usepackage{enumitem}
\setlist[itemize]{
    topsep=0pt,
    itemsep=0pt, 
    leftmargin=1.2em
}

\definecolor{findingBody}{HTML}{F2F6F9}
\definecolor{findingTitle}{HTML}{DCE7EF}
\definecolor{findingBorder}{HTML}{A5BACB}

\newcounter{finding}
\newtcolorbox{finding}[1]{
  enhanced, breakable,
  colback=findingBody, colframe=findingBorder, boxrule=0.5pt, arc=1pt,
  left=5pt, right=5pt, top=3pt, bottom=3pt,
  fontupper=\small,
  before upper={\refstepcounter{finding}\textbf{Finding~\thefinding:}~\ignorespaces}
}

\newcommand{\wy}[1]{\textcolor{red}{}}

\usepackage{CJKutf8}

\title{From Faulty Claims to Incorrect Patches: An Empirical Study of Task-Grounded Reasoning in Coding Agents}

\author{You Wang}
\orcid{0009-0009-2789-364X}
\affiliation{%
  \institution{Zhejiang University}
  \department{College of Computer Science and Technology and the State Key Laboratory of Blockchain and Data Security}
  \city{Hangzhou}
  \country{China}
}
\email{prinzywang@zju.edu.cn}

\author{Jinjun Huang}
\orcid{0009-0003-0098-2331}
\affiliation{%
  \institution{Zhejiang University}
  \department{College of Computer Science and Technology and the State Key Laboratory of Blockchain and Data Security}
  \city{Hangzhou}
  \country{China}
}
\email{huanghuanghuang@zju.edu.cn}

\author{Zhonghao Jiang}
\orcid{0009-0005-4774-2225}
\affiliation{%
  \institution{Zhejiang University}
  \department{College of Computer Science and Technology and the State Key Laboratory of Blockchain and Data Security}
  \city{Hangzhou}
  \country{China}
}
\email{zhonghao.j@zju.edu.cn}

\author{Jialun Cao}
\orcid{0000-0003-4892-6294}
\affiliation{%
  \institution{The Hong Kong University of Science and Technology}
  \department{Department of Computer Science and Engineering}
  \city{Hong Kong}
  \country{Hong Kong}
}
\email{jialuncao@ust.hk}

\author{Jiachi Chen}
\orcid{0000-0002-0192-9992}
\affiliation{%
  \institution{Zhejiang University}
  \department{College of Computer Science and Technology and the State Key Laboratory of Blockchain and Data Security}
  \city{Hangzhou}
  \country{China}
}
\email{chenjiachi@zju.edu.cn}

\author{Zhongxin Liu}
\correspondingauthor
\orcid{0000-0002-1981-1626}
\affiliation{%
  \institution{Zhejiang University}
  \department{College of Computer Science and Technology and the State Key Laboratory of Blockchain and Data Security}
  \city{Hangzhou}
  \country{China}
}
\email{liu\_zx@zju.edu.cn}

\begin{document}

\begin{abstract}
  Large language model (LLM)-based coding agents increasingly resolve real-world software engineering tasks, yet they still fail on many of them.
  Prior trajectory studies characterize these failures through task-agnostic patterns of agent behavior, such as repeated actions.
  However, incorrect patches can also stem from task-specific errors: an agent may follow a sound workflow yet act on a faulty understanding of the existing code, the task requirements, or a candidate fix. 
  Reasoning models, which power most state-of-the-art coding agents, reveal their understanding through their native chain-of-thought (CoT).
  The CoT thus serves as an artifact for analyzing whether agents understand the task correctly, by checking whether its claims about the task are consistent with the facts.
  \wy{这里就是要这样写：我们研究的对象是 reasoning，这里要说明我们为什么要研究 reasoning（要研究 task understanding --> 要研究 reasoning）。研究 task understanding 的必要性已经澄清了。task understanding 只是研究动机，不是研究内容或者研究对象。我们的标题中只有 reasoning，没有 understanding.}
  Yet it remains unclear how many such checkable claims the CoT contains, how many of them are faulty, and whether faulty claims actually cause incorrect patches.
  We present the first empirical study of such claims, which we term task-grounded reasoning, analyzing 400 trajectories of four open-weight reasoning models on SWE-bench Pro.
  We extract atomic claims from the CoT with an LLM, assess them against evidence gathered by an LLM agent, and manually trace the defects of incorrect patches back to faulty claims, confirming each link with a claim-corrected replay.
  Our results show that task-grounded reasoning covers only 18.5\% of the CoT tokens, yet yields 191 atomic claims per trajectory.
  At least 9.6\% of these claims are faulty.
  Faulty claims also occur in 96.7\% of the resolved trajectories, so their mere presence is a weak signal of failure.
  Yet two thirds of the 188 incorrect patches trace back to faulty claims.
  Correcting such a claim makes the agent fix the linked defect in 89.0\% of the cases, and correcting all such claims resolves 11.7\% of the failed tasks.
  Our findings show that the CoT carries task-specific signals for diagnosing and repairing agent failures, but exploiting them requires pinpointing the few faulty claims that actually cause patch defects.
\end{abstract}

\maketitle

\section{Introduction}

Coding agents are increasingly capable of resolving real-world software engineering tasks\cite{yang2024sweagentagentcomputerinterfacesenable,wang2025openhandsopenplatformai,jiang2025agentic}
and now solve a large share of the issues in issue-solving benchmarks~\cite{jimenez2024swe,deng2025swe,huang2026deepswe}.
Yet they still fail on many of them, often silently, submitting plausible-looking patches that pass their own validation but do not implement the intended behavior~\cite{wang2026solved}.
Improving agents therefore requires understanding not only how often they fail, but how their issue resolution goes wrong.

To understand how agents fail, prior studies analyze their trajectories through properties such as action patterns and the semantic coherence among thoughts, actions, and observations~\cite{bouzenia2025understanding,codeagentbehavior2026}, identifying failure patterns such as repeated actions or untested fixes.
Such patterns are \emph{task-agnostic}: they describe how an agent behaves, independent of what the task requires.
Yet an agent may follow an exemplary workflow free of such patterns
and still produce an incorrect patch by acting on a faulty \emph{task understanding}, such as an incorrect belief about how the existing code behaves, what the task requires, or how the requirements can be realized.
Detecting such failures requires checking this understanding against the task itself.

An agent's task understanding is not directly observable, but reasoning models, which power most state-of-the-art coding agents~\cite{anthropic2025claudecode, openai2025codex}, expose an observable proxy: their native chain of thought (CoT).
Unlike the brief thoughts that agent frameworks record alongside actions~\cite{chen2026beyond,xiao2026reducing}, the CoT contains extensive analysis of how the agent interprets the issue, what it believes the code does, and why it chooses a particular fix.
Prior work shows that models tend to externalize their reasoning in the CoT on tasks that require multi-step reasoning~\cite{korbak2025chain}, and that the CoT reveals substantially more about agent behavior than actions alone~\cite{baker2025monitoring,monitoringmonitorability2025}.
The CoT records what the agent explicitly states about the task, making it a natural artifact for studying task understanding.

\begin{figure}[t]
  \centering
  \includegraphics[width=0.97\linewidth]{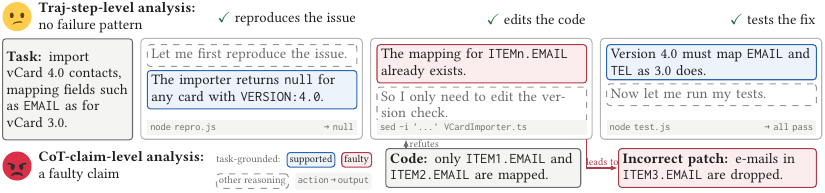}
  \caption{No task-agnostic failure pattern occurs, yet a faulty claim in the CoT leads to an incorrect patch.}
  \label{fig:intro-example}
  \Description{Three steps of a trajectory whose CoT contains supported, faulty, and other reasoning; a faulty claim refuted by the code leads to an incorrect patch.}
\end{figure}

To study task understanding, we focus on the claims the CoT makes about the task: how the original implementation behaves, where it falls short of the task requirements, what the task specifically requires, and how a fix can meet those requirements.
We refer to such content as \emph{task-grounded reasoning}.
Its defining property is that it is \emph{checkable}: whether a claim holds can be determined from available and authoritative evidence, such as the task statement, the repository at its base commit, and the documentation of the libraries it uses.
Identifying errors in this reasoning can thus be framed as fact verification~\cite{thorne2018fever,pagnoni2021frank}: we decompose task-grounded reasoning into atomic \emph{reasoning units}, each stating one claim, and call a unit refuted by evidence a \emph{faulty claim}.
For example, the agent in Figure~\ref{fig:intro-example} exhibits none of the task-agnostic failure patterns, yet submits an incorrect patch because its CoT claims that the code already maps every \texttt{ITEMn.EMAIL} field, whereas the code maps only the first two.
Faulty claims thus enable task-specific diagnosis: a faulty claim pinpoints \emph{what} the agent understands incorrectly, even when its behavior shows no sign of failure.

However, whether faulty claims can help explain why agents produce incorrect patches remains an open question.
First, it is unclear how much of the CoT meets the definition of task-grounded reasoning and how often faulty claims occur in it, which determines whether the CoT offers a basis for diagnosis at all.
Second, even if they do occur, it is unclear whether they are actually linked to a patch's defects, since CoT is not necessarily a faithful account of what a model computes~\cite{turpin2023language,arcuschin2025chain}.
Answering these questions reveals whether the extensive reasoning that agents produce can help explain and repair their failures.

This paper presents the first empirical study of task-grounded reasoning in coding agents.
We analyze 400 trajectories of four open-weight reasoning models from two model families, running in mini-swe-agent on 100 instances of SWE-bench Pro, and address three research questions:

\begin{itemize}
  \item \textbf{RQ1:} How much of the reasoning of coding agents is task-grounded, and which aspects of the task do its claims describe?
  \item \textbf{RQ2:} How common are faulty claims, what do they get wrong, and how do they relate to the outcome of a task?
  \item \textbf{RQ3:} How many incorrect patches trace back to faulty claims, and does correcting these claims repair the patches?
\end{itemize}

\noindent
For RQ1, an LLM extracts the reasoning units from the CoT of each step and classifies them into four types.
For RQ2, an LLM-based assessment agent gathers evidence from the original repository and library documentation to judge each unit as supported, refuted, or unresolved. 
The refuted units are the faulty claims.
We manually check samples of both steps to estimate their accuracy. 
For RQ3, we locate the concrete defects of each incorrect patch via corrective edits that make it pass the validation tests while meeting the task requirements, search for faulty claims in the reasoning that these defects trace back to, and replay the trajectory with such a claim corrected right before submission to check whether the agent fixes the linked defect.

Our study leads to the following key findings:
\begin{itemize}
  \item Task-grounded reasoning covers only 18.5\% of the CoT tokens, yet amounts to 191 checkable claims per trajectory.
    Newer models make about twice as many claims as their predecessors.
  \item At least 9.6\% of the claims are faulty, and 97.8\% of the trajectories contain at least one.
    Agents understand the task better than the code: 5.5\% of the claims about the requirements are faulty, versus 13.5\% about the original implementation and 14.7\% about fix mechanisms.
    Rather than referring to nonexistent APIs or code, most faulty claims misstate how existing code behaves.
  \item The number of faulty claims is a weak indicator of failure: 96.7\% of the resolved trajectories also contain faulty claims, which often do no harm because the agent recovers from them or its patch does not depend on them.
    Yet some faulty claims leave defects that the validation tests miss, yielding plausible but incorrect patches.
  \item Two-thirds of the incorrect patches (127 of 188) trace back to faulty claims, yet these claims account for only about one in 26 faulty claims in the failed trajectories.
    Correcting such a claim causes the agent to fix the linked defect in 89.0\% of cases with a small revision, though only 11.5\% resolve the task, as other defects remain or the tests miss the fixed behavior.
\end{itemize}
These findings show that faulty claims are a substantial cause of incorrect patches, but their mere presence is a weak signal of failure.
Precise diagnosis of failed trajectories thus calls for failure attribution tools that establish accurate causal links between faulty claims and patch defects.

In summary, this paper contributes the following:
\begin{itemize}
  \item \textbf{Methodology.} A claim-level methodology that combines LLM-based extraction and assessment of task-grounded reasoning with a manual tracing of patch defects to faulty claims, whose links are confirmed by claim-corrected trajectory replays.
  \item \textbf{Empirical study.} The first empirical study of task-grounded reasoning and its faulty claims across four reasoning models, from their prevalence to the patch defects they cause.
  \item \textbf{Implications.} Insights for using the reasoning of coding agents to diagnose, prioritize, and repair their failures, both at runtime and during training.
\end{itemize}

\section{Methodology}

\subsection{Data Collection}
\label{sec:data-collection}

We run mini-swe-agent
~\cite{yang2024sweagentagentcomputerinterfacesenable} with four open-weight reasoning models, DeepSeek-V3.2~\cite{liu2025deepseek}, DeepSeek-V4.1-Flash~\cite{xu2026deepseek}, Qwen3.6-27B~\cite{qwen3.6-27b}, and Qwen3.8-27B~\cite{qwen38}, on SWE-bench Pro~\cite{deng2025swe}, and record the native CoT of each step along with its action and observation.
Unlike prior trajectory studies, which compare different agents~\cite{bouzenia2025understanding}, we fix the scaffold and vary the model, since the CoT is produced by the model and differences in reasoning are expected to arise mainly between models.
We consider only open-weight models because commercial models typically do not expose their native CoT~\cite{openai2026gpt6astra, anthropic2026claudefable5}.
We set the thinking budget of all models to 16,384 tokens per step to allow sufficient reasoning before each action.
We sample 100 instances on which all four models terminate normally and submit a patch.
This yields 400 trajectories with 28,001 steps in total (Table~\ref{tab:dataset}).

\begin{table}[t]
  \caption{Overview of the collected trajectories.}
  \label{tab:dataset}
  \centering
  \fontsize{8.5pt}{10.5pt}\selectfont
  \setlength{\tabcolsep}{5pt}
  \scalebox{0.95}{
  \begin{tabular}{@{}lcccc@{}}
    \toprule
    \textbf{Model} & \textbf{Release Date} & \textbf{Resolved} & \textbf{Total Steps} & \textbf{Total Reasoning Tokens} \\
    \midrule
    DeepSeek-V3.2       & 2025-12-01 & 43/100 (43.0\%) & 8,834 & 1,219,334 \\
    DeepSeek-V4.1-Flash & 2026-09-10 & 64/100 (64.0\%) & 6,825 & 4,616,470 \\
    Qwen3.6-27B         & 2026-04-22 & 44/100 (44.0\%) & 5,440 &   887,232 \\
    Qwen3.8-27B         & 2026-08-14 & 61/100 (61.0\%) & 6,902 & 3,190,453 \\
    \bottomrule
  \end{tabular}
  }
\end{table}

\subsection{Task-Grounded Reasoning and Faulty Claims}
\label{sec:concepts}

Our analysis targets the native CoT that the model generates at each step of a trajectory.
In this section, we first define which part of the CoT we study and the unit in which we analyze it, then classify these units into four types, and finally define the faulty claims among them.

\begin{definition}[Task-grounded reasoning]
\emph{Task-grounded reasoning} is the part of the CoT that makes task-relevant claims whose truth can be objectively determined by evidence from authoritative sources such as the task statement, the original repository, and external knowledge (e.g., the API of a library).
\end{definition}
In addition to task-grounded reasoning, the CoT also plans edits, navigates the repository, and tracks the agent's progress.
Such content does not make checkable claims about the task and is therefore outside the scope of our study.

We operationalize this definition as follows: a claim is task-grounded if it concerns the task to be solved and can be judged true or false without reference to the edits the agent has made.
This criterion concerns whether a claim is checkable, not whether it is correct, so mistaken beliefs about the repository, misreadings of the task, and tentative conclusions all qualify.
It excludes content without a substantive claim, such as editing intentions, open questions, navigation, and test or submission status.
It also excludes claims about the agent's own implementation, whose truth depends on the agent's edits.

\begin{definition}[Task-grounded reasoning unit]
A \emph{task-grounded reasoning unit} is the smallest self-contained proposition in task-grounded reasoning that can be independently judged true or false, retaining the details that the CoT attaches to it, such as its subject, conditions, and scope.
\end{definition}
For brevity, we refer to task-grounded reasoning units as \emph{reasoning units} in the remainder of the paper.
A reasoning unit need not coincide with a sentence: one sentence may contain several propositions, and one proposition may span several clauses.

\begin{table}[t]
  \caption{Types of reasoning units.}
  \label{tab:types}
  \centering
  \fontsize{8.5pt}{10.5pt}\selectfont
  \begin{tabular}{@{}>{\centering\arraybackslash}m{0.24\linewidth}p{0.72\linewidth}@{}}
    \toprule
    \textbf{Type} & \textbf{Definition / Example} \\
    \midrule
    \underline{O}riginal implementation\par (O) &
      \begin{tabular}[c]{@{}p{\linewidth}@{}}
        \textbf{Definition:} A claim about how the original implementation, including the libraries it uses, behaves or is structured. \\
        \textbf{Example:} ``\texttt{argparse}’s \texttt{type} parameter expects a callable that converts a string into a value of the desired type.''
      \end{tabular} \\
    \midrule
    Implementation \underline{G}ap\par (G) &
      \begin{tabular}[c]{@{}p{\linewidth}@{}}
        \textbf{Definition:} A claim about where the original implementation falls short of the task requirements, such as missing functionality or the code responsible for a bug. \\
        \textbf{Example:} ``The current \texttt{run} method discards the wrapped function’s result, although it must return that result for both synchronous and asynchronous calls.''
      \end{tabular} \\
    \midrule
    \underline{S}emantic constraint\par (S) &
      \begin{tabular}[c]{@{}p{\linewidth}@{}}
        \textbf{Definition:} A claim about what the task requires, such as the expected behavior, inputs and outputs, interfaces, or compatibility. \\
        \textbf{Example:} ``\texttt{CommandParser} must raise \texttt{EmptyCommandError} when no command is provided.''
      \end{tabular} \\
    \midrule
    \underline{F}ix mechanism\par (F) &
      \begin{tabular}[c]{@{}p{\linewidth}@{}}
        \textbf{Definition:} A claim that a concrete mechanism, such as an algorithm or a data representation, satisfies a requirement or closes a gap. \\
        \textbf{Example:} ``Iterating through matching patterns in reverse insertion order gives precedence to the most recently added value.''
      \end{tabular} \\
    \bottomrule
  \end{tabular}
\end{table}

We classify each reasoning unit into one of four types according to its role in solving the task, as summarized in Table~\ref{tab:types}.
Type O describes how the original implementation behaves, such as what a function returns or how a module handles a given input.
Type G describes where the original implementation falls short of the task requirements, such as which upstream functionality a requested feature still lacks or which code causes a reported bug; G claims are thus closely tied to locating the code to change.
Type S describes what the task requires, such as expected behavior, interfaces, or compatibility constraints that a correct solution must satisfy.
Type F claims that a concrete mechanism, such as an algorithm or a data representation, satisfies a requirement or closes a gap.
Unlike an editing intention, which only states what the agent will do, an F claim asserts what a mechanism achieves, and can thus be checked against the requirement it addresses.

The task-grounded reasoning in the CoT is not always correct.
We define: 

\begin{definition}[Task-grounded faulty claim]
A \emph{task-grounded faulty claim} is a reasoning unit that is contradicted by evidence from the authoritative sources that determine the truth of task-grounded reasoning, concerning the same subject under the same conditions and scope.
\end{definition}

For brevity, we refer to task-grounded faulty claims as \emph{faulty claims}.
A claim is faulty only if the evidence contradicts it. 
A claim that the available sources neither confirm nor refute is not considered faulty.
Faulty claims can occur in all four types, for example, an incorrect statement about the structure or behavior of the repository code or a library it uses, an incorrect localization or explanation of a bug, a misunderstanding of the specific behavior that the task requires, or a mechanism that does not achieve what it is claimed to achieve.

\subsection{Extracting and Classifying Task-Grounded Reasoning}
\label{sec:rq1-method}

To address RQ1, we extract the reasoning units from the CoT for each step across the 400 trajectories and classify each unit into one of the four types.
Because manually annotating all 28,001 steps is infeasible, we automate extraction and classification with an LLM and then manually check a sample of its output.

We use GPT-5.6-Luna\cite{openai2026gpt56luna}, with the reasoning effort set to \texttt{xhigh}, and process each step in a separate request.
The prompt contains the definitions of Section~\ref{sec:concepts} and two classification examples for each type. 
Besides the complete CoT of the step, each request contains the task statement and the observations received since the previous step, which help the LLM resolve references to code and behavior.
We do not provide the oracle patch or the test results, so that the LLM does not label claims in hindsight.
Within a single request, the LLM identifies the task-grounded claims in the CoT, extracts them as units, and assigns each unit a type.
For each unit, the LLM outputs a normalized statement of the claim, a verbatim quote from the CoT that supports it, its type, and any files it refers to.
We check each quote against the CoT; if a quote does not match, we send the request again, up to one more time.

To validate the automated results, two authors first independently annotate the same 10 steps and then discuss their disagreements to align their understanding of the definitions and types.
Each of them then checks 150 randomly sampled steps with at least one extracted unit, 300 distinct steps in total, for units that do not meet the definition of a reasoning unit and units with incorrect types.
We report the results of this check alongside those of RQ1.

\subsection{Identifying Faulty Claims}
\label{sec:rq2-method}

To address RQ2, we assess each reasoning unit extracted in RQ1 and identify the faulty claims within it.
Unlike classification, this assessment requires comparison with evidence that the CoT does not contain.
We therefore automate the assessment with an LLM agent that gathers the evidence itself.
To distinguish it from the coding agents under study, we call it the \emph{assessment agent}.

We build the assessment agent upon mini-swe-agent, using GPT-5.6-Luna as the underlying model, with the reasoning effort set to \texttt{max}.
Each run assesses all units of one type from one step. 
For each unit, the agent receives its normalized claim, its quote, and the files it refers to, together with the task statement and the complete CoT for the step, which help the agent interpret the claims.
The agent works in a fresh container with the original repository at the base commit of the task.
It does not receive the observations of the trajectory, the generated patch, the oracle patch, or the test results, so that its judgment rests on the evidence it gathers rather than on the coding agent's edits or the outcome of the trajectory.
Four tools are available: a shell for inspecting the repository and running tests, a fetcher for the official documentation and upstream source of libraries, a citation tool that records the exact source excerpts the assessment relies on, and a submission tool.
Besides the definition of faulty claims, the prompt gives instructions specific to each type:
\begin{itemize}
  \item \textbf{O:} The agent traces the claimed structure or behavior through the relevant code.
  \item \textbf{G:} The agent compares the relevant original implementation with the requirement and checks whether it falls short where and how the claim states.
  \item \textbf{S:} The agent reconstructs the requirement from the task statement and looks for an input where the claimed requirement and the actual one yield different outcomes.
  \item \textbf{F:} The agent checks whether the mechanism achieves its claimed effect by looking for a counterexample under the conditions of the claim.
\end{itemize}

For each unit, the agent compares the claim with the evidence it has gathered and, following the three verdicts of fact verification in FEVER~\cite{thorne2018fever}, judges the claim as \emph{supported}, \emph{refuted}, or \emph{unresolved} (NotEnoughInfo in FEVER) if the evidence neither confirms nor refutes it.
A unit is a faulty claim only if it is refuted.
Every judgment comes with an explanation, and refuted and supported judgments must also cite evidence.
We check that the submission contains exactly one assessment per unit and that every cited excerpt has been recorded; otherwise, the agent receives feedback and resubmits within the same run.

To validate the automated results, we randomly sample 10 of the 100 instances and take the trajectories of all four models on them, 40 trajectories in total.
Two authors first independently annotate the faulty claims in the same two trajectories and then discuss their disagreements to align their understanding of the definition of faulty claims and the error types.
Each of them then checks all faulty claims identified in the trajectories of 5 instances for whether they are indeed faulty.
They further classify each confirmed faulty claim into one of the four error types in Table~\ref{tab:error-types} according to how it is wrong.

\begin{table}[t]
  \caption{Error types of faulty claims.}
  \label{tab:error-types}
  \centering
  \fontsize{8.5pt}{10.5pt}\selectfont
  \begin{tabular}{@{}>{\centering\arraybackslash}m{0.24\linewidth}p{0.72\linewidth}@{}}
    \toprule
    \textbf{Error Type} & \textbf{Definition / Example} \\
    \midrule
    Misreference &
      \begin{tabular}[c]{@{}p{\linewidth}@{}}
        \textbf{Definition:} The claim refers to an object that does not exist or is not the intended one, such as a wrong identifier, file, endpoint, or code location. \\
        \textbf{Example:} ``The go-redis/v9 \texttt{Options} struct includes \texttt{MaxConnIdleTime} for configuring the maximum idle time for connections.'' The field is named \texttt{ConnMaxIdleTime}.
      \end{tabular} \\
    \midrule
    Misstatement &
      \begin{tabular}[c]{@{}p{\linewidth}@{}}
        \textbf{Definition:} The claim refers to the right object in the right situation but misstates what the object does or must do. \\
        \textbf{Example:} ``The \texttt{at()} result for an index of \texttt{NaN} is \texttt{0}.'' \texttt{NaN} is converted to index~0, so \texttt{at()} returns the first element.
      \end{tabular} \\
    \midrule
    Omission &
      \begin{tabular}[c]{@{}p{\linewidth}@{}}
        \textbf{Definition:} The claim holds for the cases it covers but leaves out other applicable cases, facts, or required work. \\
        \textbf{Example:} ``The \texttt{cockroachdb} implementation uses \texttt{crdb} only for \texttt{Lock}.'' \texttt{SetVersion} also calls \texttt{crdb.ExecuteTx}.
      \end{tabular} \\
    \midrule
    Overgeneralization &
      \begin{tabular}[c]{@{}p{\linewidth}@{}}
        \textbf{Definition:} The claim holds only under some conditions but is stated for a broader scope. \\
        \textbf{Example:} ``All quirks in \texttt{\_Quirk} use the generated name and check \texttt{quirk.name not in skip}.'' The Google Docs quirk sets the explicit name \texttt{ua-googledocs} instead.
      \end{tabular} \\
    \bottomrule
  \end{tabular}
\end{table}

We do not check for missed faulty claims, because confirming every reasoning unit would require prohibitive manual effort.
The precision measured in this check thus yields a lower-bound estimate of the number of faulty claims.
We report the results of this check alongside those of RQ2.

\subsection{Tracing Patch Defects to Faulty Claims}
\label{sec:rq3-method}

To address RQ3, we start with the incorrect patches in the unresolved trajectories, identify their concrete defects, and trace each defect back to faulty claims in the reasoning.
We then correct these claims during the trajectory and check whether the patch improves accordingly.
We divide the unresolved trajectories between two authors, who analyze them as follows.

\textit{Locating patch defects.}
For each unresolved trajectory, each of the two authors examines the incorrect patch and the report of the failed tests, and constructs an edit that, applied on top of the incorrect patch, makes it pass all validation tests while meeting the specific requirements of the task statement.
We call this edit the \emph{corrective patch}; it captures why the incorrect patch fails.

\textit{Identifying key faulty claims.}
Guided by the corrective patch, each author reads through the complete CoT of the trajectory and identifies the reasoning units in it that are faulty claims related to a defect in the patch.
A faulty claim is related to a defect if, had the claim been corrected, the agent would have had a chance to generate a patch without that defect.
We call such claims \emph{key faulty claims}.
Because the tasks in SWE-bench Pro are complex, the generated patches often span hundreds of lines \cite{deng2025swe}, and an incorrect patch may contain multiple defects.
A key faulty claim is thus usually related to only some of them and accounts for only part of the corrective patch. 
We record this part as the \emph{targeted fix} of the claim.
We search all reasoning units rather than only the faulty claims identified in RQ2 because the automated assessment may miss faulty claims, and this step aims for the highest possible recall.
Manual effort remains affordable because we only look for faulty claims related to patch defects, rather than all faulty claims in the trajectory.
This also allows us to report the recall of the automated assessment in RQ2 on the key faulty claims.

\textit{Correcting key faulty claims.}
To confirm the link between a key faulty claim and the defect, we replay the unresolved trajectory up to the point right before its submission, send a user message that corrects the claim, and ask the agent to revise its patch accordingly and submit it again.
We then compute the difference between the new patch and the original one.
Each of the two authors who identified the claim compares this difference with the targeted fix and assesses whether it contains a change that fixes the same defect as the targeted fix, possibly implemented differently.
If so, we consider the correction of the claim to fix the corresponding defect.
Unlike prior work that replays a trajectory from the step where an error occurs~\cite{codetracer,wang2026trajaudit}, we replay it up to the submission.
The step of a faulty claim usually precedes the submission, often before the original patch is complete, so replaying from there may yield a final patch that differs too much from the original one to be compared with the targeted fix.
Besides checking the targeted fixes, we run the validation tests on each new patch and report how many unresolved trajectories pass more validation tests or become resolved.
We also perform one replay per trajectory that corrects all its key faulty claims in a single user message; for this replay, we report only these test outcomes.

\section{Results}

This section presents the results for the three research questions.
We abbreviate Qwen3.6-27B and Qwen3.8-27B as Qwen3.6 and Qwen3.8, and task-grounded reasoning as TGR in figures.

\subsection{RQ1: Task-Grounded Reasoning in the CoT}
\label{sec:rq1}

We first study how much of the CoT is task-grounded reasoning and which types of claims it contains.
From the 400 trajectories, the extraction of Section~\ref{sec:rq1-method} yields 76,537 reasoning units.
On the 10 steps (143 units) that both authors annotate independently, their agreement is high (Cohen's $\kappa = 0.937$).
Of the 2,149 units in the 300 manually checked steps, 2,006 (93.3\%) are correct; 62 (2.9\%) are not reasoning units, and 81 (3.8\%) have an incorrect type.
To measure how much of the CoT text is task-grounded, we tokenize the CoT with the \texttt{o200k\_base} tokenizer~\cite{openai_tiktoken} and count a token as task-grounded if it is covered by the quote of at least one unit.

\subsubsection{Prevalence of Task-Grounded Reasoning}

\leavevmode\par

\begin{figure}[t]
  \centering
  \includegraphics[width=\linewidth]{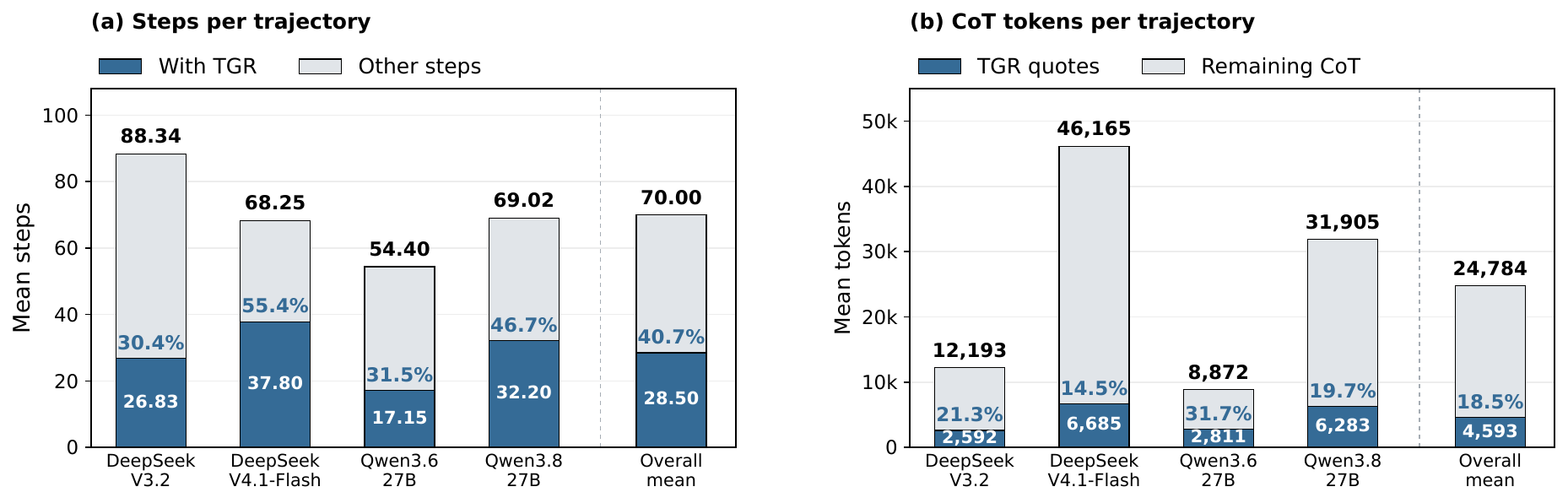}
  \caption{Mean steps (a) and CoT tokens (b) per trajectory for each model and over all 400 trajectories, with the part containing task-grounded reasoning; percentages give its share of the bar.}
  \label{fig:rq1-prevalence}
\end{figure}

\textit{Task-grounded reasoning makes up less than a fifth of the CoT.}
Figure~\ref{fig:rq1-prevalence} shows the number of trajectory steps that contain task-grounded reasoning and the number of CoT tokens they cover.
On average, a trajectory has 70.0 steps, 28.5 (40.7\%) of which contain at least one reasoning unit, and 24,784 CoT tokens, of which the quotes of reasoning units cover 4,593 (18.5\%).
The agent thus makes checkable claims about the task in two out of five steps, but these claims make up less than a fifth of what it writes; the rest of the CoT plans edits, navigates the repository, and tracks progress.
Yet this small fraction contains 191.3 reasoning units per trajectory, 6.7 per step with task-grounded reasoning.

\textit{Newer models think longer, but task-grounded reasoning grows less.}
Within each model family, the newer model, which also resolves more tasks (Table~\ref{tab:dataset}), thinks much longer: DeepSeek-V4.1-Flash writes 3.79 times as many CoT tokens as DeepSeek-V3.2, and Qwen3.8 3.60 times as many as Qwen3.6.
Task-grounded reasoning grows far less, by a factor of 2.58 and 2.24 in tokens and 2.11 and 1.88 in units, respectively.
Put differently, only 12.0\% of the additional CoT tokens of DeepSeek-V4.1-Flash and 15.1\% of those of Qwen3.8 are task-grounded.
The longer reasoning of the newer models therefore contains more claims about the task, but it is mostly devoted to other content, and task-grounded reasoning becomes sparser within it.

\begin{finding}{Prevalence of Task-Grounded Reasoning}
  Task-grounded reasoning covers only 18.5\% of the CoT tokens, yet amounts to 191 checkable claims per trajectory.
  Newer models think 3.6--3.8$\times$ longer but make only about twice as many claims, so CoT length overstates how much an agent reasons about the task.
\end{finding}

\subsubsection{Types of Reasoning Units}

\leavevmode\par

\begin{figure}[t]
  \centering
  \includegraphics[width=\linewidth]{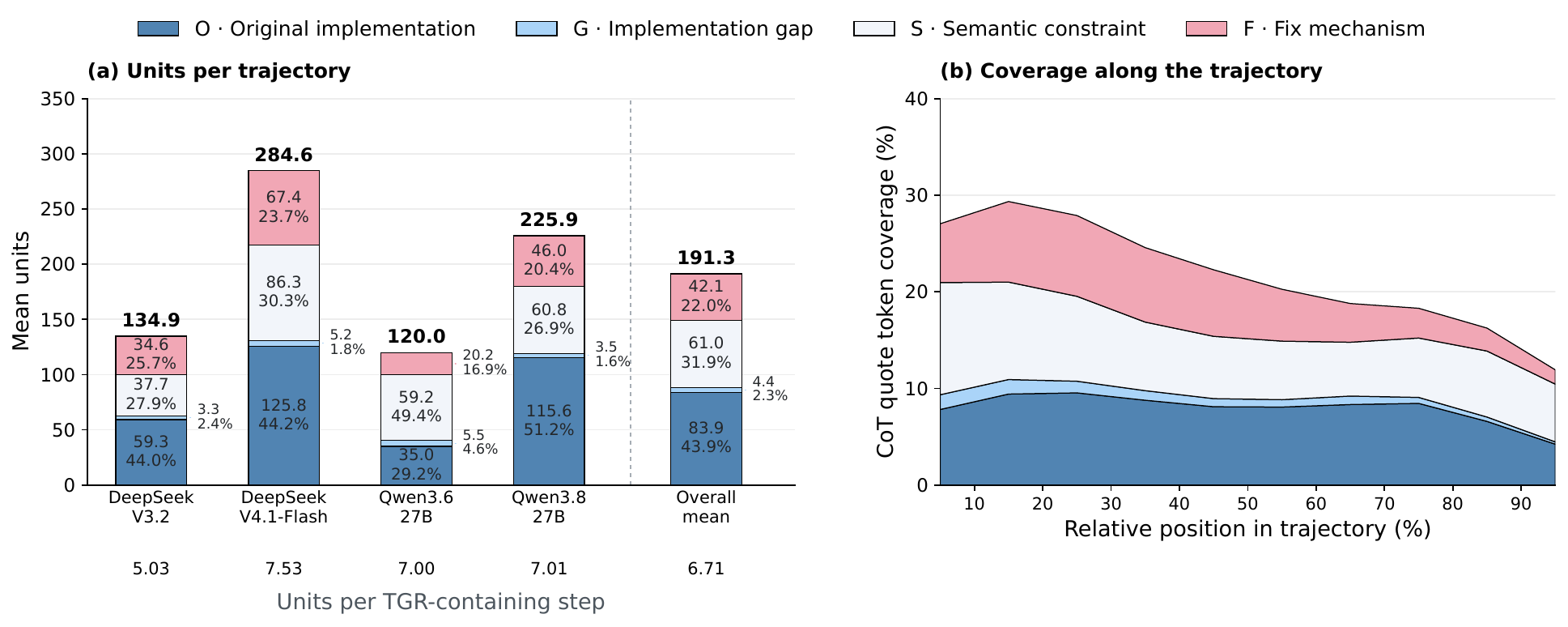}
  \caption{Mean reasoning units per trajectory by type (a) and token coverage of each type along the trajectory, divided into ten bins by relative step position and averaged over the four models (b).}
  \label{fig:rq1-types}
\end{figure}

\textit{Task-grounded reasoning mostly describes the code and the requirements.}
Figure~\ref{fig:rq1-types}(a) breaks the reasoning units down by type.
Across all trajectories, O accounts for 43.9\% of the units, S for 31.9\%, F for 22.0\%, and G for only 2.3\%.
About three-quarters of the claims thus describe either how the original implementation behaves or what the task requires, whereas G claims are rare.
This is expected, because a gap is the conclusion of comparing the original implementation with the requirements: a task typically has only a few such conclusions, whereas reaching and acting on each of them draws on many facts about the code and many details of the required behavior.

\textit{Newer models spend their additional claims differently.}
The composition of the claims also shows what the additional reasoning of the newer models is about.
Qwen3.8 makes 105.9 more units per trajectory than Qwen3.6, 80.6 of which are O and 25.8 F, whereas the number of S units remains virtually unchanged (59.2 vs.\ 60.8).
The drop in the S share from 49.4\% to 26.9\% does not mean that Qwen3.8 states fewer requirements; it states as many, but reasons about how the existing code behaves more than three times as much.
DeepSeek-V4.1-Flash, in contrast, produces 149.8 more units per trajectory than DeepSeek-V3.2, and the increase is distributed across O (+66.5), S (+48.6), and F (+32.8).
Each of these types grows about twofold, so the composition of DeepSeek-V4.1-Flash's claims hardly differs from that of DeepSeek-V3.2 (e.g., 44.2\% vs.\ 44.0\% O).

\textit{Mechanisms come early, requirements return at the end.}
Figure~\ref{fig:rq1-types}(b) shows how the types are distributed along a trajectory.
Task-grounded reasoning is front-loaded: its token coverage peaks at 29.3\% in the second tenth of the trajectory and declines to 11.9\% in the last.
Surprisingly, F claims are concentrated at the beginning rather than the end of a trajectory: in every model, 61.3--77.0\% of the F units lie in the first half of the trajectory, compared with 51.9--65.2\% of the units of all types, and 30--49 of the 100 trajectories already contain an F claim in their first tenth.
Rather than working out a mechanism after exploring the code, agents sometimes take it from the task statement. 
For example, in \texttt{navidrome-0488fb9}, Qwen3.6 states in step~1 of 63, before running any command, that it will ``update \texttt{Key()} method to include the \texttt{square} parameter to prevent cache collisions'', closely following the requirement that cache keys include the new parameter.
The rest of the trajectory is then spent implementing and checking such mechanisms.
O claims, in contrast, stay at about 8--10\% of the CoT tokens until 80\% of the trajectory, as agents keep inspecting the code, and S claims, although they also decline, make up half of the task-grounded tokens in the last tenth (6.0\% of 11.9\%), when agents check their solution against the requirements before submitting it.

\begin{finding}{Types of Reasoning Units}
  O and S make up three-quarters of the claims, while G, the conclusion drawn from them, only 2.3\%.
  Qwen3.8 spends 76\% of its additional units on the original implementation, whereas DeepSeek-V4.1-Flash makes about twice as many claims across all types.
  Surprisingly, mechanisms are proposed early, sometimes taken from the task statement before the code is explored, whereas requirements make up half of the task-grounded reasoning at the end.
\end{finding}

\subsection{RQ2: Faulty Claims in Task-Grounded Reasoning}
\label{sec:rq2}

We next study how many reasoning units are faulty claims, what these claims get wrong, and how they relate to the outcome of a trajectory.
The assessment agent of Section~\ref{sec:rq2-method} judges all 76,537 units: it supports 62,587 (81.8\%), refutes 8,540 (11.2\%), and neither supports nor refutes the remaining 5,410 (7.1\%).
We refer to the refuted units as the faulty claims and define the error rate of a group of units
 as the number of faulty claims divided by the number of units in the group.
On the 45 faulty claims in the two trajectories that both authors annotate independently, their agreement is high (Cohen's $\kappa = 0.877$).
Of the 896 faulty claims identified by the assessment agent across the 40 sampled trajectories, our manual check (Section~\ref{sec:rq2-method}) confirms 773, yielding a precision of 86.3\%.

\subsubsection{Prevalence of Faulty Claims}

\leavevmode\par

\begin{figure}[t]
  \centering
  \includegraphics[width=0.92\linewidth]{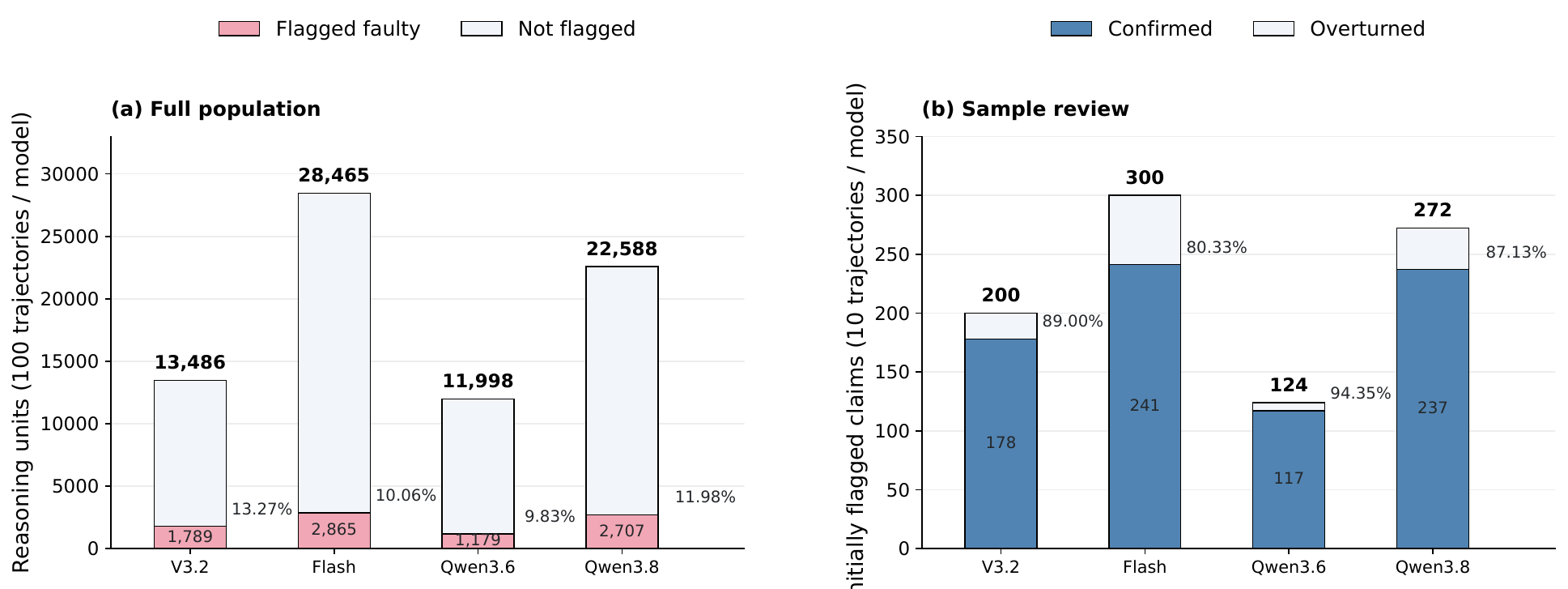}
  \caption{Faulty claims among all reasoning units (a) and their manual check in the 10 sampled trajectories per model (b). Percentages give the error rate (a) and the precision (b). V3.2 and Flash denote DeepSeek-V3.2 and DeepSeek-V4.1-Flash.}
  \label{fig:rq2-prevalence}
\end{figure}

\textit{At least one in ten claims is faulty.}
The assessment agent refutes 8,540 of the 76,537 reasoning units (11.2\%, Figure~\ref{fig:rq2-prevalence}(a)).
Discounting this rate by the precision of the manual check yields an estimated error rate of 9.6\%; as we do not check for missed faulty claims, this is a lower-bound estimate. 
Faulty claims occur in 391 of the 400 trajectories (97.8\%): every trajectory of DeepSeek-V3.2 contains at least one, and even Qwen3.6, the model with the fewest faulty claims, has them in 93 of its 100 trajectories.
A trajectory contains 21.4 faulty claims on average (median 14).
Faulty claims are therefore not confined to a few derailed runs, but a regular part of how the agents reason about a task.
The manual check in Figure~\ref{fig:rq2-prevalence}(b) shows that this prevalence does not stem from false positives of the assessment agent: each of the 40 sampled trajectories contains confirmed faulty claims, which make up 9.4\% of their 8,215 units, close to the estimated 9.6\%.

\textit{Stronger models do not make fewer faulty claims.}
The newer model of each family resolves considerably more tasks (Table~\ref{tab:dataset}), yet it makes more faulty claims than its predecessor.
DeepSeek-V4.1-Flash has a lower error rate than DeepSeek-V3.2 (10.1\% vs.\ 13.3\%), but because it makes 2.11 times as many claims (Section~\ref{sec:rq1}), it ends up with 60\% more faulty claims (2,865 vs.\ 1,789).
Qwen3.8 does not even have a lower error rate: it makes 1.88 times as many claims as Qwen3.6 but 2.30 times as many faulty claims, and its error rate rises from 9.8\% to 12.0\%.
The longer reasoning of the newer models observed in RQ1 thus contains not only more claims about the task, but also more wrong ones.
More faulty claims, however, do not necessarily indicate less reliable reasoning.
Rather, the newer models may reason more deeply about the task, e.g., about how the code behaves on more complex inputs, and such questions are harder to answer correctly.
Qwen3.8, for example, spends most of its additional claims on the original implementation (Section~\ref{sec:rq1}), one of the most error-prone types (Section~\ref{sec:rq2-types}).
This shift alone accounts for most of the rise in its error rate from 9.8\% to 12.0\%: even if each type were as often wrong as for Qwen3.6, Qwen3.8 would reach 11.7\%.

\begin{finding}{Prevalence of Faulty Claims}
  At least about one in ten reasoning units (an estimated 9.6\%) is a faulty claim, 97.8\% of the trajectories contain at least one, and the manual check confirms 86.3\% of the sampled faulty claims.
  The newer models resolve more tasks but make 60\% (DeepSeek) and 130\% (Qwen) more faulty claims than their predecessors, possibly because they reason more deeply and thus about harder questions, rather than because their reasoning is less reliable.
\end{finding}

\subsubsection{What Faulty Claims Get Wrong}\label{sec:rq2-types}

\begin{figure}[t]
  \centering
  \includegraphics[width=\linewidth]{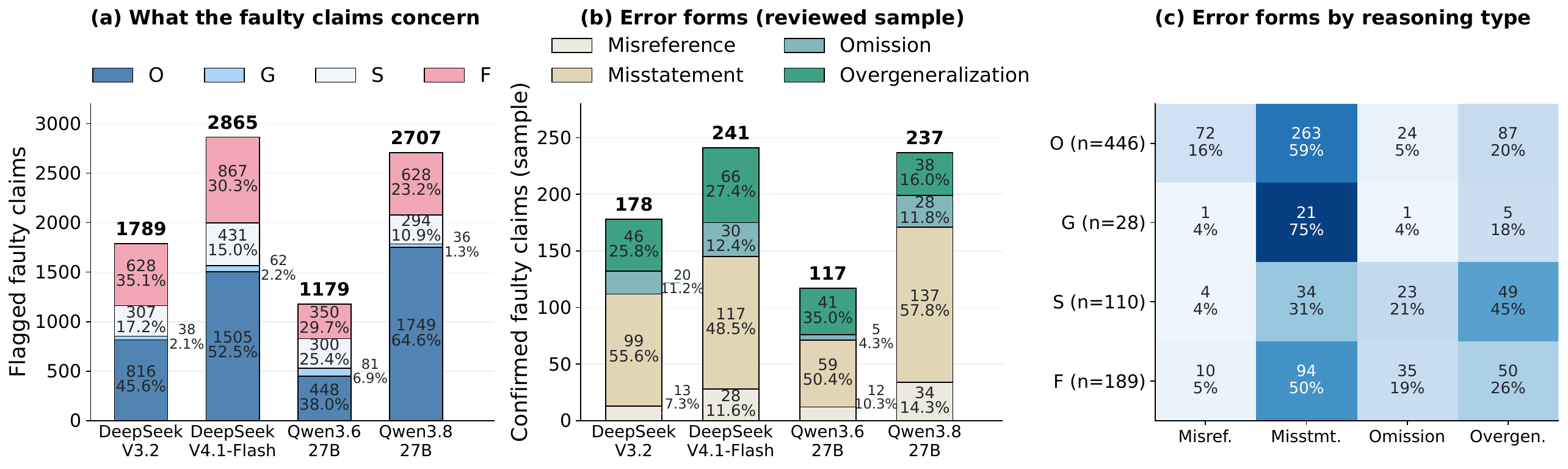}
  \caption{Faulty claims by type among all reasoning units (a), confirmed faulty claims by error type in the sampled trajectories (b), and error types of the confirmed faulty claims of each type (c). Percentages are relative to the confirmed faulty claims of a model (b) or of a type (c).}
  \label{fig:rq2-types}
\end{figure}

\leavevmode\par

\textit{Agents understand the task better than the code.}
Figure~\ref{fig:rq2-types}(a) shows that four in five faulty claims concern the original implementation (O) or a fix mechanism (F).
This is not only because the agents make many such claims: O and F claims are also more than twice as likely to be wrong as S claims (13.5\% and 14.7\% vs.\ 5.5\%).
S claims have the lowest error rate in every model (4.8--8.1\%, vs.\ at least 10.2\% for the other types) and are thus the most reliable type.
The difference lies in where the answer comes from.
An S claim can usually be read off the task statement, which in SWE-bench Pro spells out the required behavior in detail.
An O claim, in contrast, must be inferred from code that the agent has read only in part, and an F claim predicts the behavior of code that does not even exist yet.
The agents thus rarely misunderstand what they are asked to do, but often misunderstand how the code works and what their changes will do.

\textit{Agents misunderstand real objects rather than invent them.}
Figure~\ref{fig:rq2-types}(b) shows how the confirmed faulty claims are wrong.
Misstatements are by far the most common error type (53.3\%), followed by Overgeneralizations (24.7\%), whereas Misreferences (11.3\%) and Omissions (10.7\%) are rare.  
Misstatement is the largest error type for every model.
Only about one in nine faulty claims thus refers to a wrong or nonexistent object, the kind of error commonly associated with hallucination in code generation~\cite{liu2024exploring}.
Working in the repository, the agents can look up identifiers and mostly name the right functions, fields, and files; what they get wrong is how these objects behave and interact. 
Such errors are harder to spot than a wrong name: every object that the claim mentions exists, so the claim looks plausible, and only tracing the code reveals that it is wrong.

\textit{Requirements are wrong in scope, code in behavior.}
Figure~\ref{fig:rq2-types}(c) shows how a claim is wrong depends on what it is about.
By far the most common combination is a claim about the original implementation that misstates what the code does: it accounts for 59.0\% of the O errors and for a third of all confirmed faulty claims.
Such a claim can silently undermine a patch that is otherwise correct.
In \texttt{flipt-492cc0b}, DeepSeek-V3.2 adds a \texttt{validate} method to the Redis configuration and claims that ``the config loading will call validate on \texttt{RedisCacheConfig} because it has a validate method''.
The loader, however, only validates the top-level configuration sections, so the method is never called, and the patch accepts the invalid Redis settings that the task requires it to reject.
Faulty requirements, in contrast, are mostly wrong in scope: two-thirds of the S errors are Overgeneralizations or Omissions, i.e., the agent understands what a requirement demands, but applies it to more or fewer cases than the task intends.
Faulty mechanisms show both kinds of error in about equal parts.

\begin{finding}{What Faulty Claims Get Wrong}
  Agents understand the task better than the code: 13.5\% of the O and 14.7\% of the F claims are faulty, whereas only 5.5\% of the S claims are faulty.
  Only 11.3\% of the confirmed faulty claims refer to a wrong object; most get real objects wrong. The most common error, a third of all confirmed faulty claims, misstates how the original implementation behaves.
  Faulty requirements are mostly wrong in scope.
\end{finding}

\subsubsection{Faulty Claims and Task Outcomes}\label{sec:rq2-outcomes}

\begin{figure}[t]
  \centering
  \includegraphics[width=0.8\linewidth]{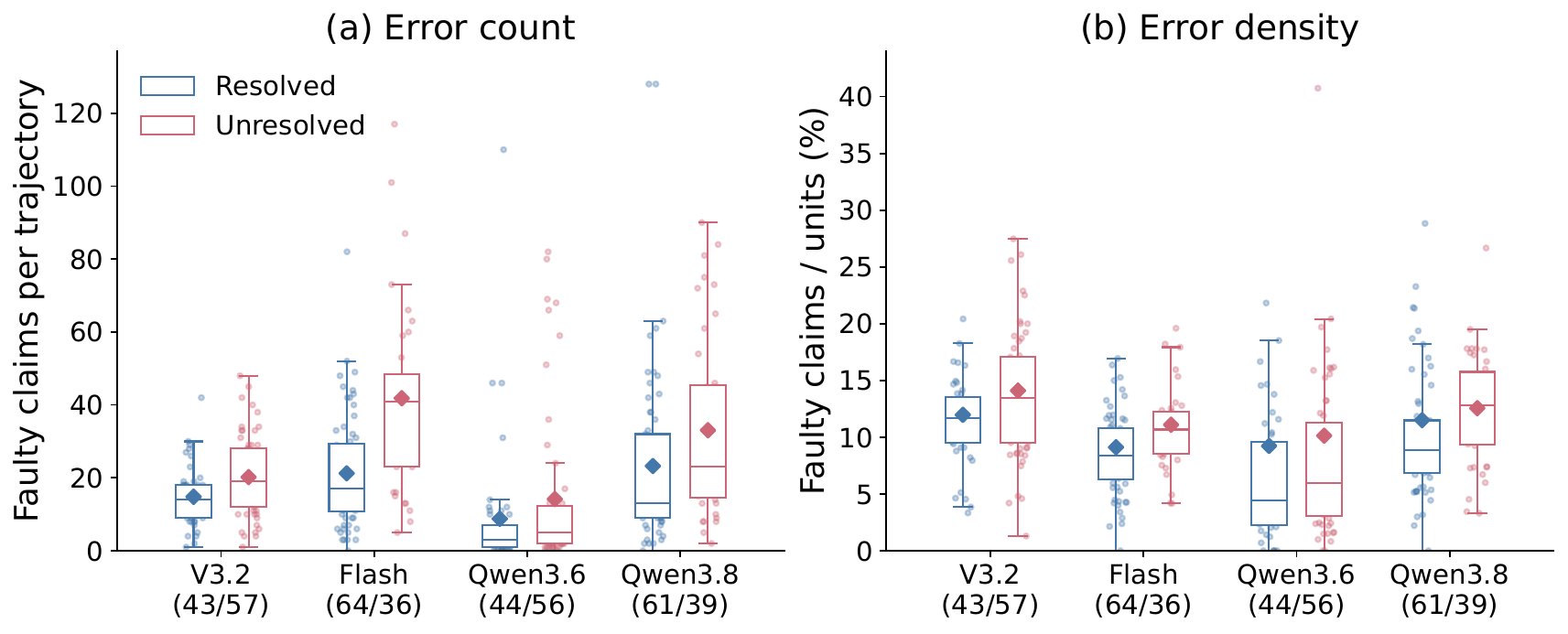}
  \caption{Faulty claims (a) and error rate (b) per trajectory in resolved and unresolved trajectories. Diamonds give the mean (a) and the error rate over all units of the group (b); numbers below the models give the resolved/unresolved trajectories.}
  \label{fig:rq2-outcomes}
\end{figure}

\leavevmode\par

\textit{Unresolved trajectories contain more faulty claims, but only a slightly higher error rate.}
Figure~\ref{fig:rq2-outcomes}(a) shows that in all four models, unresolved trajectories contain more faulty claims than resolved ones (25.2 vs.\ 17.9 on average), mostly because they contain more claims overall: for DeepSeek-V4.1-Flash, for example, they contain 1.97 times as many faulty claims, but also 1.62 times as many units.
Each claim is thus only slightly more likely to be wrong in an unresolved trajectory (Figure~\ref{fig:rq2-outcomes}(b)): the error rate is higher in every model, but only by 0.9--2.1 percentage points, and on the 38 tasks that some models resolve and others do not, by 1.9 percentage points on average, which is not statistically significant (95\% bootstrap interval $-0.2$ to $4.0$).
The number of faulty claims in a trajectory is therefore a weak indicator of failure, as it largely reflects how much the agent reasons.

\textit{Not all faulty claims lead to incorrect patches.}
Faulty claims are common even in resolved trajectories: 205 of the 212 (96.7\%) contain at least one, and the two trajectories with the most faulty claims in our data (128 each) both resolve their tasks.
We identify two reasons why a faulty claim may not lead to an incorrect patch: 

\noindent\textit{(1) The agent recovers from the claim.}
Evidence that the agent obtains later, from the repository or from execution, may contradict the claim before the patch builds on it.
In \texttt{element-web-a692fe2}, DeepSeek-V3.2 claims that \texttt{cli.\allowbreak getCapabilities()} is ``likely what we need to check server-side policy'', but reading the existing tests in the next step, it finds the dedicated method and uses that instead.
In \texttt{qutebrowser-46e6839}, Qwen3.8 wrongly assumes that Qt's \texttt{QVersionNumber} pads missing segments with zeros, so that \texttt{5.4} equals \texttt{5.4.0}, but when a check of its own fails, it corrects the assumption.
Recovery prevents the defect, but may cost several steps.

\noindent\textit{(2) The patch does not depend on the claim.}
A claim may be wrong only about code that the patch does not rely on.
In \texttt{qutebrowser-5e0d6dc}, DeepSeek-V4.1-Flash claims that ``all quirks in \texttt{\_Quirk} use generated name'', which the existing Google Docs quirk contradicts. 
The new quirk it adds does use a generated name, however, so the patch handles it correctly.

\textit{Resolved patches may still be incorrect.}
Conversely, a resolved trajectory does not guarantee that its faulty claims are harmless.
A faulty claim may introduce a defect that the validation tests do not cover, so that the agent submits a \emph{plausible} patch, i.e., one that passes the tests but is incorrect~\cite{wang2026solved}.
In \texttt{qutebrowser-a84ecfb}, Qwen3.8 reasons that ``the requirements only ask that \texttt{CommandParser} accepts the argument'' and leaves its command suggestions disabled by default; the missing behavior is covered only by an end-to-end test that the evaluation harness does not run.
In \texttt{qutebrowser-5e0d6dc}, DeepSeek-V4.1-Flash keeps an assignment in its \texttt{Array.prototype.at} polyfill that makes the method visible to \texttt{for...in} loops over arrays, because ``tests won't check''.
In both cases, the agent narrows the task to what it expects to be tested, and the faulty claim behind the defect is stated explicitly in the reasoning, although the test results do not reveal it.

\begin{finding}{Faulty Claims and Task Outcomes}
  The number of faulty claims is a weak indicator of failure: unresolved trajectories contain more faulty claims mainly because they contain more claims, not because their claims are more often wrong.
  96.7\% of the resolved trajectories also contain faulty claims, which often do no harm because the agent later recovers from them or the patch does not rely on them.
  Yet some faulty claims leave defects that the evaluation tests miss, yielding plausible but incorrect patches.
\end{finding}

\subsection{RQ3: From Faulty Claims to Patch Defects}
\label{sec:rq3}

Finally, we study how many patches that fail the validation tests trace back to faulty claims and whether correcting these claims repairs the patches.
Following Section~\ref{sec:rq3-method}, we analyze the 188 unresolved trajectories, whose failing tests let us locate the defects of their patches, and replay them with their key faulty claims corrected.

\begin{table}[t]
  \caption{Outcomes of correcting key faulty claims, per unresolved trajectory with all its key faulty claims corrected (left) and per key faulty claim corrected separately (right). Percentages are relative to the first column of each half. V3.2 and Flash denote DeepSeek-V3.2 and DeepSeek-V4.1-Flash.}
  \label{tab:rq3-overview}
  \centering
  \fontsize{8pt}{10pt}\selectfont
  \setlength{\tabcolsep}{2.7pt}
  \scalebox{0.95}{
  \begin{tabular}{@{}lcccc|cccc@{}}
    \toprule
    & \multicolumn{4}{c|}{\textbf{Unresolved trajectories}} & \multicolumn{4}{c}{\textbf{Key faulty claims}} \\
    \textbf{Model} & \textbf{Total} & \textbf{w/ key claim} & \textbf{More tests passed} & \textbf{Resolved} & \textbf{Total} & \textbf{Defect fixed} & \textbf{More tests passed} & \textbf{Resolved} \\
    \midrule
    V3.2    & 57  & 44 (77.2\%)  & 13 (22.8\%) & 9 (15.8\%)  & 61  & 47 (77.0\%)  & 11 (18.0\%) & 8 (13.1\%) \\
    Flash   & 36  & 24 (66.7\%)  & 5 (13.9\%)  & 4 (11.1\%)  & 34  & 34 (100\%)   & 5 (14.7\%)  & 4 (11.8\%) \\
    Qwen3.6 & 56  & 32 (57.1\%)  & 12 (21.4\%) & 6 (10.7\%)  & 47  & 41 (87.2\%)  & 11 (23.4\%) & 6 (12.8\%) \\
    Qwen3.8 & 39  & 27 (69.2\%)  & 7 (17.9\%)  & 3 (7.7\%)   & 40  & 40 (100\%)   & 7 (17.5\%)  & 3 (7.5\%) \\
    \midrule
    Total   & 188 & 127 (67.6\%) & 37 (19.7\%) & 22 (11.7\%) & 182 & 162 (89.0\%) & 34 (18.7\%) & 21 (11.5\%) \\
    \bottomrule
  \end{tabular}
  }
\end{table}

\subsubsection{Prevalence of Key Faulty Claims}

\leavevmode\par

\textit{Two thirds of the failing patches trace back to key faulty claims.}
Table~\ref{tab:rq3-overview} shows that 127 of the 188 unresolved trajectories (67.6\%) contain at least one key faulty claim, i.e., a faulty claim whose correction gives the agent a chance to avoid a defect in its patch.
The share ranges from 57.1\% for Qwen3.6 to 77.2\% for DeepSeek-V3.2.
Most of these trajectories contain a single key faulty claim (80), while 47 contain two to four, for a total of 182 key faulty claims.
In the remaining 61 unresolved trajectories, we find no faulty claim directly linked to a patch defect, although such claims may still contribute to the failure indirectly.

\textit{Few faulty claims are directly linked to patch defects.}
Although a large proportion of the failing patches trace back to faulty claims, only a small fraction of the faulty claims directly lead to a defect.
The assessment agent of RQ2 identifies 4,741 faulty claims in the unresolved trajectories, 25.2 per trajectory, but only 182 are key, about one per trajectory.
The manually checked sample of RQ2 shows the same ratio: of the 452 confirmed faulty claims in its 20 unresolved trajectories, only 17 (3.8\%) are key.
The assessment agent refutes 17 of the 20 key faulty claims in these trajectories, so it detects most of them, but buries each among about 26 other faulty claims that are not directly linked to a defect.
To verify an agent's reasoning, detecting faulty claims is only half the problem; the other half is deciding which of them deserve the agent's attention.

\begin{figure}[t]
  \centering
  \includegraphics[width=0.95\linewidth]{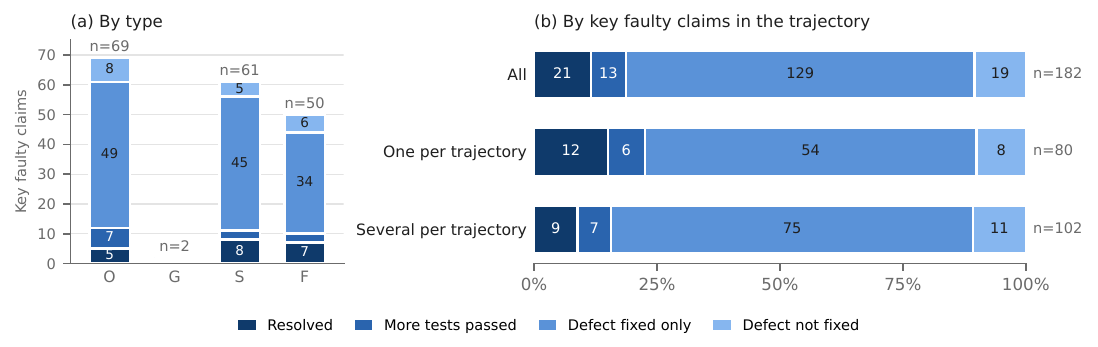}
  \caption{Outcome of correcting a single key faulty claim, as counts by the type of the claim (a) and as shares by the number of key faulty claims in its trajectory (b). Each tier excludes the tiers before it.}
  \label{fig:rq3-outcomes}
\end{figure}

\textit{Faulty claims about the code, the requirements, and the fix all contribute to patch defects.}
Of the key faulty claims, 69 (37.9\%) are O claims, 61 (33.5\%) S, 50 (27.5\%) F, and only 2 G (Figure~\ref{fig:rq3-outcomes}(a)).
O claims assume facts about the code and its dependencies that do not hold, e.g., that an API or a dependency version provides a feature, or that the existing code already handles part of the task.
In \texttt{navidrome-d21932b}, for instance, DeepSeek-V3.2 assumes that the check \texttt{isWritable} guards all playlist updates, but it guards only changes to the tracks, so the patch still lets other users overwrite the metadata of a public playlist.
Many F claims take a mechanism that covers part of the required behavior as sufficient.
In \texttt{teleport-87a5935}, DeepSeek-V4.1-Flash detects an unreachable SQL Server by the substring ``connection refused'' and concludes that ``the string check is sufficient'', but the driver reports timeouts as ``unable to open tcp connection with host'', which the check misses.
Although RQ2 found S claims to be the most reliable type, misread requirements lie behind a third of the defects that we trace to faulty claims.
In \texttt{NodeBB-0e07f3c}, for example, DeepSeek-V4.1-Flash must reserve system tags for privileged users and decides in step~28: ``Should system tags be case-sensitive? Probably exact.''
The repository, however, lowercases and trims every tag before storing it, so an ordinary user can still create the reserved tag \texttt{locked} by entering \texttt{LOCKED}.
A tentative guess about the requirement thus becomes a permission bypass in the patch.

\begin{finding}{Prevalence of Key Faulty Claims}
  Two-thirds of the failing patches (127/188) trace back to at least one key faulty claim, yet only about one in 26 faulty claims in unresolved trajectories is key.
  Key faulty claims concern the original implementation (37.9\%), the requirements (33.5\%), and the fix mechanism (27.5\%); requirements thus account for a third, although agents get them wrong less often than any other type of claim.
\end{finding}

\subsubsection{Correcting Key Faulty Claims}

\begin{figure}[t]
  \centering
  \includegraphics[width=0.95\linewidth]{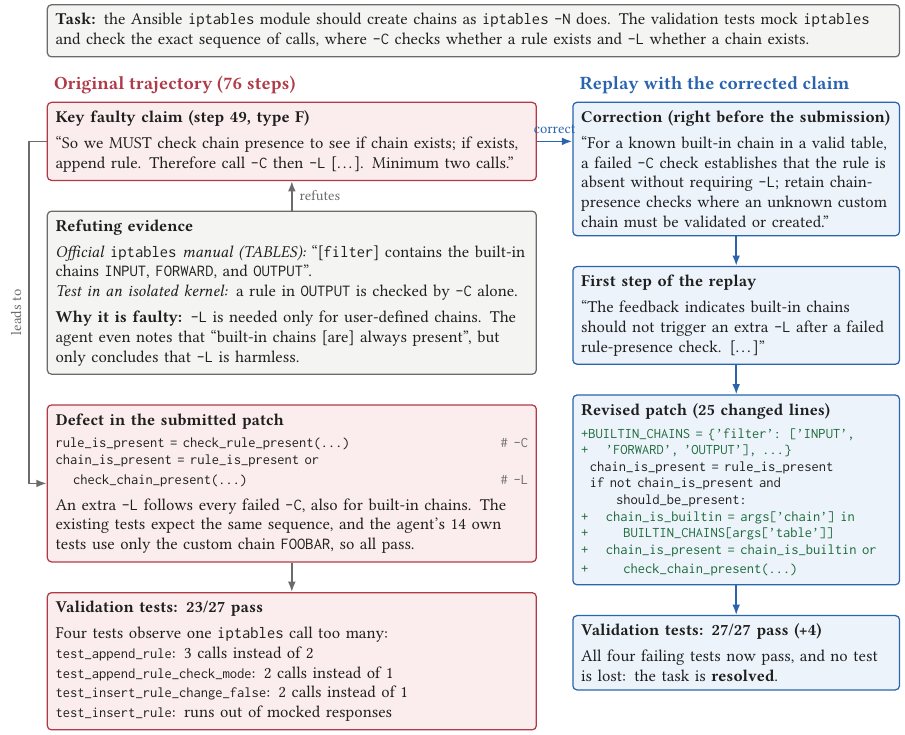}
  \caption{A key faulty claim of DeepSeek-V4.1-Flash in \texttt{ansible-a1569ea} (left) and the replay with the corrected claim (right). Code is abridged.}
  \label{fig:rq3-example}
\end{figure}

\leavevmode\par
\textit{An example.}
Figure~\ref{fig:rq3-example} traces one key faulty claim from the trajectory to the resolved task.
In \texttt{ansible-a1569ea}, DeepSeek-V4.1-Flash overgeneralizes that every failed \texttt{iptables} rule check must be followed by a chain-existence check, when this is only necessary for user-defined chains.
The assessment agent of RQ2 refutes it with evidence from two sources: the \texttt{iptables} manual, whose \texttt{filter} table lists \texttt{INPUT}, \texttt{FORWARD}, and \texttt{OUTPUT} as built-in chains, and a test in an isolated kernel showing that \texttt{-C} alone suffices for a rule in \texttt{OUTPUT}.
Our correction, sent right before submission, explains the distinction. 
The agent then revises its patch to skip the redundant check for built-in chains and resolves the task.

\textit{Test results undercount the repaired defects.}
Figure~\ref{fig:rq3-outcomes}(b) summarizes the outcomes of correcting each of the 182 key faulty claims separately.
In 162 of these replays (89.0\%), the revised patch contains the targeted fix of the corrected claim, i.e., it fixes the defect linked to the claim.
Yet only 34 (18.7\%) pass more validation tests, and 21 (11.5\%) resolve the task.
In the other 129 replays that fix the defect, the fix does not show up in the test results.
One reason is that the tests do not exercise the fixed behavior, as for the permission bypass in \texttt{NodeBB-0e07f3c}.
Another reason is that other defects remain in the patch.
In the 47 trajectories with several key faulty claims, 91 of the 102 single corrections fix the linked defect, but only 9 (8.8\%) resolve the task, compared with 12 of 80 (15.0\%) in trajectories with a single key faulty claim.
This also shows that when all key faulty claims of these trajectories are corrected in one message, 19 of the 47 trajectories (40.4\%) then pass more tests and 10 (21.3\%) are resolved, whereas single corrections do so in only 16 and 9 of themin these trajectories are corrected in a single message, 19 of the 47 trajectories (40.4\%) pass more tests and 10 (21.3\%) are resolved, whereas single corrections do so for only 16 and 9 trajectories.
That single corrections already resolve some of these trajectories suggests, again, that the tests miss the defects of their other key faulty claims.

\textit{Agents act on corrections with small, focused edits.}
To produce the targeted fix, the agents do not rewrite their patches: after the correction, they take a median of 17 further steps, and the revised patch differs from the original one in a median of 13 lines, whereas the original patches change a median of 101 lines.
The key faulty claims thus do not merely co-occur with the defects; they cause them, since the updated patch changes where and how the corrected claim says it should.
The newer models follow corrections more reliably: DeepSeek-V4.1-Flash and Qwen3.8 produce the targeted fix in every replay, DeepSeek-V3.2 in 77.0\% and Qwen3.6 in 87.2\%.
In about half of the 20 replays without the targeted fix, the agent leaves the defective code untouched, and in three of them, all by DeepSeek-V3.2, it resubmits the original patch unchanged.
In the others, it acts on the correction but implements it incorrectly.
In \texttt{openlibrary-09865f5}, for instance, Qwen3.6 is told that the Infogami \texttt{Thing} class is serialized with its \texttt{dict()} method rather than a nonexistent \texttt{to\_dict()}; it replaces the call to \texttt{to\_dict()} with \texttt{dict(obj)}, swapping one wrong API for another.

\textit{What remains unresolved.}
Of the 188 unresolved trajectories, 127 (67.6\%) contain at least one key faulty claim.
Correcting all their key faulty claims makes 37 of the 188 trajectories (19.7\%) pass more validation tests and resolves 22 (11.7\%) more (Table~\ref{tab:rq3-overview}).
Over the 100 tasks, this would raise the resolution rate from 53.0\% to 58.5\% on average across the four models.
Most failing trajectories thus remain unresolved even after all their key faulty claims are corrected.
One reason is that key faulty claims account for only part of the defects: in the 127 unresolved trajectories with such claims, their targeted fixes make up only 40.1\% of the lines changed by the corrective patch on average.
The other 61 unresolved trajectories contain no key faulty claim at all.
Some defects have no counterpart in the reasoning and arise only when the agent edits the code.
In \texttt{navidrome-66b74c81}, Qwen3.6 adds a new file to its patch by writing the diff by hand, but the hunk header declares 77 lines while the hunk contains 81, so the submitted file is truncated and the build fails, although the file compiled in the agent's environment.
In other cases, the agent reasons correctly but implements its conclusion incompletely.
In \texttt{openlibrary-25858f9f}, DeepSeek-V3.2 moves the Solr helpers of \texttt{update\_work.py} into a new module and lists \texttt{solr\_update} among the functions to move; it deletes the function from the old module and imports it from the new one, but never adds it there, so both test modules fail to import.
Such defects leave no faulty claim in the reasoning.
Nevertheless, key faulty claims appear to be a worthwhile target: they lie behind two-thirds of the failing patches, and a single message that corrects one of them fixes the linked defect in 89.0\% of the replays with a small revision.
Faulty claims thus explain a substantial part of agent failures, though not all; other defects appear to be introduced during editing, whether the reasoning overlooks them or is correct.

\begin{finding}{Correcting Key Faulty Claims}
  Correcting a key faulty claim in a single message before submission makes the agent fix the linked defect in 89.0\% of the cases with a median revision of 13 lines, confirming the causal link.
  Yet only 18.7\% of the corrections pass more validation tests and 11.5\% resolve the task, as other defects remain or the tests miss the fixed behavior.
  Correcting all such claims resolves 22 of the 188 unresolved trajectories: their targeted fixes make up only 40.1\% of the lines changed by the corrective patch on average, and the remaining defects likely stem from edits that go beyond or depart from the reasoning.
\end{finding}

\section{Discussion}

\subsection{Implications}

\hspace*{1em}\textit{Implications for Researchers.}
\textbf{(1) Reasoning carries task-specific signals of failure.}
Where it is accessible, as for open-weight models, the native CoT can serve as an artifact for diagnosing, monitoring, and repairing coding agents.
It could also supervise training: verified claims could provide process rewards denser than test outcomes. 
Exploiting the CoT also requires more than detecting faulty claims: since most never reach the patch, reporting all of them buries the few that matter.
Signals for prioritization include whether the final patch still builds on a claim and whether later evidence revises it.
Our key faulty claims, linked to their defects, can serve as ground truth for such prioritization.
\textbf{(2) Cover what the reasoning leaves out.}
In a third of the failing patches, no defect can be traced back to a faulty claim in the CoT.
Some of their defects lie in edits that the reasoning never discusses, 
and some in edits that contradict its reasoning. 
Verifying stated claims cannot catch such defects.
Future work can check the consistency between reasoning and edits, extending thought--action alignment~\cite{bouzenia2025understanding} from single actions to the patch: tracing each hunk back to the reasoning that motivates it flags undisclosed edits, and tracing each conclusion forward to the patch flags edits that contradict or omit it.
\textbf{(3) Faulty claims profile model weaknesses.}
Resolution rates show that an agent fails, but not which of its beliefs are wrong.
Faulty claims reveal a consistent profile across models: agents most often misstate how existing code behaves or what their changes will do, and misread requirements less often, yet mostly in scope and behind a third of the key faulty claims.
Post-training could thus target repository-level code comprehension, e.g., predicting the behavior of partially read code and the effect of an edit, and the scope of requirements, rather than only hallucinated APIs.

\textit{Implications for Practitioners.}
\textbf{(1) Verify the claims a patch relies on.}
When using models whose CoT is visible, practitioners can verify the claims behind the hunks of a patch.
This verification need not happen inside the agent: it can serve as a checkpoint before submission that returns a correction for each refuted claim to the agent, or as a post-processing step after submission, when users review a patch or when tools detect overfitting patches.
Most faulty claims refer to the right objects but misstate their behavior or scope, so looking up identifiers is not enough.
Instead, the verification should check that the code behaves as the agent describes it (e.g., with execution probes), that a proposed fix achieves the required behavior (e.g., with targeted tests), and that a stated requirement matches the task statement.
\textbf{(2) Capture anti-patterns in the reasoning.}
Some statements in the reasoning reveal where a patch may go wrong.
Hedged claims such as ``Probably exact'' mark what the agent is uncertain about but never checks, and statements such as ``tests won't check'' reveal shortcuts that narrow the task to what the agent expects to be tested.
Capturing such anti-patterns can flag patches for closer inspection, even when they pass all tests.

\subsection{Limitations and Threats to Validity}

Our study has several limitations.
First, we do not check the recall of the assessment agent, since manually confirming all 76,537 reasoning units is prohibitive; the error rates of faulty claims are thus lower bounds.
Yet the assessment agent refutes 150 of the 182 key faulty claims (82.4\%) identified in RQ3, so it likely misses few of the faulty claims that matter for the patch.
Second, due to the manual effort, we validate the automated extraction and assessment only on random samples of 300 steps and 40 trajectories. 
However, these samples cover all four models and exceed the sizes required for a 99\% confidence level with a 5\% margin of error, with 2,149 of the 76,537 units (660 required) and 896 of the 8,540 faulty claims (618 required).
Third, we study only open-weight models in a single scaffold and benchmark, which may limit generalizability.
However, commercial models typically hide their native CoT, whereas the CoT of open-weight models is accessible and can inform their training. 
Our models also span two families and two generations, and fixing the scaffold isolates the differences between models.
Finally, we study neither how faulty claims depend on each other nor why the models make them.
We focus on key faulty claims that are directly linked to patch defects and treat the models as black boxes, characterizing what faulty claims get wrong and how they affect the patch; both questions are left for future work.

\section{Related Work}

\hspace*{1em}\textit{Empirical studies of agent trajectories.}
Bouzenia and Pradel~\cite{bouzenia2025understanding} found that failed trajectories of software engineering agents exhibit anti-patterns such as repeated actions and untested fixes, and misalignment between thoughts and actions.
Majgaonkar et al.~\cite{codeagentbehavior2026} found that failed trajectories are longer and often still locate the right files.
Other studies model trajectories as graphs of actions to reveal strategies and inefficiencies~\cite{liu2026process}, identify behaviors that predict success~\cite{mehtiyev2026beyond}, detect structural anti-patterns such as search loops~\cite{shu2026resolve}, reveal passing trajectories of poor process quality~\cite{sahoo2026agentlens}, and trace agent workflows from localization to patch validation~\cite{ceka2026understanding}.
Closer to our work, Liu et al.~\cite{liu2025empirical} attributed most failures of agentic issue-solving tools to flawed reasoning, and Zhao et al.~\cite{zhao2026failure} found that most failures of CLI coding agents stem from epistemic errors such as acting on false premises.
Beyond agents, studies of reasoning models in single-turn code generation categorize the behaviors in their reasoning~\cite{halim2025thinking} or rate the quality of reasoning steps~\cite{xue2026reasoning}.
Unlike these studies, which characterize how agents act, annotate a root cause per trajectory or step, or assess reasoning without checking its correctness, we verify whether individual claims in the reasoning of coding agents are correct and confirm their effect on the patch.

\textit{Failure localization in agent trajectories.}
Automated failure attribution aims to pinpoint the agent and the step responsible for a failed task.
Zhang et al.~\cite{zhang2025which} formulated this problem for LLM multi-agent systems, defining the decisive error as the earliest step whose correction turns the failure into success, and found that LLMs localize fewer than 15\% of such steps.
Subsequent techniques improve attribution with spectrum-based fault localization over repeated executions~\cite{ge2026spectrum}, hierarchical causal graphs~\cite{wang2026chief}, and constraints synthesized from tool schemas and domain policies~\cite{barke2026agentrx}.
DoVer~\cite{ma2026dover} and AgentDebug~\cite{zhu2025where} further intervene at the suspected step and re-execute the agent to validate the attribution or recover from the failure.
MAST~\cite{cemri2025mast}, TRAIL~\cite{deshpande2025trail}, and TrajDebug~\cite{qi2026trajdebug} provide failure taxonomies and annotated traces, partly from coding agents.
For coding agents, CodeTracer~\cite{codetracer} reconstructs hierarchical traces to localize the earliest failure-critical step, TrajAudit~\cite{wang2026trajaudit} guides an investigator agent with test-failure output to the decisive error step, and Wink~\cite{nanda2026wink} detects misbehaviors such as specification drift and injects course-correction guidance at runtime.
These techniques localize failures to agents or steps, whereas we locate them in individual claims of the native CoT, which reveal what the agent misunderstands about the task.

\section{Conclusion}

This paper presents the first empirical study of task-grounded reasoning in coding agents, i.e., checkable claims in their native CoT about how the existing code behaves, where it falls short of the task, what the task requires, and how a fix meets those requirements.
Analyzing 400 trajectories of four open-weight reasoning models on SWE-bench Pro, we find that task-grounded reasoning covers only 18.5\% of CoT tokens, yet yields 191 claims per trajectory, at least 9.6\% of which are faulty.
Claims about code are more often faulty than those about requirements, and most faulty ones misstate how existing code behaves rather than referring to nonexistent code.
Most resolved trajectories also contain faulty claims, so their mere presence is a weak signal of failure.
Yet two-thirds of incorrect patches trace back to them, and correcting such a claim before submission makes the agent fix the linked defect in 89.0\% of cases.
Thus, the CoT carries task-specific signals for diagnosing and repairing agent failures, but exploiting them requires pinpointing the few faulty claims that patches depend on.
Future work includes failure attribution that links faulty claims to patch defects, consistency checks between reasoning and edits, and verified claims as process rewards for training.

\section{Data Availability}
Our code and data are available: \url{https://doi.org/10.5281/zenodo.23093905}.

\bibliographystyle{ACM-Reference-Format}
\bibliography{references}

\end{document}